\documentclass[prb,superscriptaddress,twocolumn]{revtex4}

\usepackage{bm}
\usepackage{graphicx}
\usepackage{amsmath,amssymb}
\usepackage{hyperref}
\usepackage[latin1]{inputenc}

\newcommand{\figwidth}{0.85\columnwidth}

\newcommand{\sign}{\operatorname{sign}}

\renewcommand{\Im}{\operatorname{Im}}

\begin{document}

\title{Transient and finite size effects in transport properties of a quantum wire}
\author{Mariano Salvay}
\author{An\'{\i}bal Iucci}
\author{Carlos Na\'on}

\affiliation{Departamento de F\'{\i}sica, Facultad de Ciencias
Exactas, Universidad Nacional de La Plata and IFLP-CONICET, CC 67,
 1900 La Plata, Argentina.}

\begin{abstract}
We study the time-dependent backscattered current produced in a
quantum wire when a local barrier is suddenly switched on. Previous
investigations are improved by taking into account the finite length
of the device. We establish two different regimes in terms of the
relationship between the energy scales associated to the voltage and
the length of the system.  We show how previous results, valid for
wires of infinite length, are modified by the finite size of the
system. In particular our study reveals a rich pattern of temporal
steps within which the current suffers an initial relaxation
followed by temporary revivals. By employing both analytical and
numerical methods we describe peculiar features of this structure.
From this analysis one concludes that our results render a recently
proposed approach to the determination of the Luttinger parameter
$K$, more realistic.
\end{abstract}


\maketitle

\section{Introduction}

Recently there has been much interest in the study of quantum
transport in quasi-one-dimensional (1D) materials, such as quantum
wires and carbon nanotubes
\cite{giamarchi04_book_1d,diventra08_book_electrical_transport,nazarov09_quantum_transport}.
In 1D systems the effect of electron-electron (e-e) correlations
cannot be disregarded, giving rise to a dramatic departure from the
Fermi liquid picture of usual condensed matter systems. This new
state of matter is called a Luttinger liquid
(LL)\cite{voit95_1D_fermi_liquids}, and is characterized by
correlation functions that decay with distance through exponents
that depend on the e-e interaction \cite{bockrath99_exponents}. This
peculiar behavior leads to the prediction of striking phenomena such
as spin-charge separation and charge fractionalization, both
experimentally
confirmed\cite{auslaender05_spin-charge_separation,jompol09_spin-charge_separation}
 \cite{steinberg08_fractionalization}. It is specially interesting to analyze time-dependent aspects of transport
\cite{cini80,pastawski92,wingreen93,arrachea02} in this context.
Indeed, the interplay between correlation-induced effects and
dynamical phenomena, originated in the out of equilibrium nature of
transport, is a highly non trivial problem that deserves much
attention. In fact, a detailed knowledge of the LL behavior in the
presence of time-dependent perturbations will facilitate the
development of devices based on quantum computation, single electron
transport and quantum interferometers\cite{fujisawa06,foa09}. But
there is also a more profound motivation to discuss this problem,
since it is directly connected to basic issues such as the evolution
of the ground state and currents after a quench, according to the
influence of different initial conditions
\cite{cazalilla06_quench_LL,perfetto10_memory_effects_transport}. In
a recent work\cite{salvay10_transient_impurity} we have analyzed the
response of a LL (formed by electrons in a quantum wire) to a sudden
switch of an interaction of the system with an external field (a
local barrier created by applying a voltage to a narrow metal gate
electrode in a single-walled carbon nanotube
\cite{biercuk04_gating_cn,biercuk05_quantum_dots_cn,biercuk05_conductance_quantization_cn}).
When such a barrier, that can be considered as a backscattering
impurity, is turned on at a finite time $t_0$, a backscattered
current $I_{bs}$ is produced. In Ref.
\onlinecite{salvay10_transient_impurity} a quantitative relation
between the time evolution of $I_{bs}$ and the constant $K$, related
to e-e correlations, was obtained. Thus, it was shown that $K$ could
be determined by measuring time intervals within the reach of
recently developed pump-probe techniques with femtosecond-attosecond
time resolutions \cite{gouielmakis07_attsecond_response}. This
result was obtained under the assumption that the length of the wire
was infinite. It is then very important, in order to make the
predictions more realistic, to understand the role that the finite
size of the sample will play in the time evolution of the electrical
current. This is the main purpose of this article. It is worth
mentioning that the effects of finite length of a quantum wire on
its transport properties were previously investigated in Refs.
\onlinecite{ponomarenko97_asymptotic_expansion},
\onlinecite{dolcini03_conductance_impurity},
\onlinecite{dolcini05_finite_length_temperature} and
\onlinecite{cheng06_transport_LL_time_dependent}, but for the case
of a static impurity. Besides its direct interest in the area of
one-dimensional strongly correlated systems, our work reveals some
general features of time-dependent transport in confined geometries
that could contribute to a deeper understanding of the behavior of
other materials, such as graphene nanoribbons under the sudden
switch-on of a constant electric field
\cite{lewkowicz09_particle-hole_dynamics_graphene} or a constant
bias voltage \cite{perfetto10_time_dependent_transport_graphene}.
Another interesting problem, closely connected with our findings, is
the role of interacting leads in transport through a quantum
point-contact. Concerning this issue it has been recently shown that
the switching process has a large impact on the relaxation and the
steady-state behavior of the current
\cite{perfetto10_memory_effects_transport}. However, for the sake of
clarity let us stress that, in contrast to the situation discussed
in Ref.\onlinecite{perfetto10_memory_effects_transport}, where the
interaction between the wire and the contacts that leads to the bias
voltage V is also switched on at a given time, here we only consider
the switching of the time-dependent impurity, whereas the external
voltage is assumed to act at all times. Our study is also of
interest in the context of cold atomic gases, where quantum quenches
are being intensively investigated
\cite{greiner02_fast_tunnability,cazalilla06_quench_LL,kollath07_quench_BH,cramer08_relaxation_superlattices,iucci09_quench_LL,iucci10_quench_sine_gordon,trotzky11_quench_local_relaxation}.

The paper is organized as follows. In Section II we present the
model and define the backscattered current $I_{bs}$. In Section III,
in order to facilitate the understanding of our main results,
concerned with finite size effects, we recall the results obtained
for an infinite wire. Section IV contains the original contributions
of this paper. We present a closed expression that gives the
time-dependence of the backscattered current as an integral of a
function that contains, through an infinite product, the effect of
the finite length $L$ of the wire and the position of the impurity.
By combining both analytical and numerical methods, we examine the
evolution of the current for different wire sizes (Figs. 2-4).
Finally, in Section V, we summarize our results and conclusions.

\section{The model}

We consider a clean Luttinger liquid of length $L$ adiabatically
coupled to two electrodes at its end points ($x = \pm L/2$) with
chemical potentials $\mu_{L}$ and $\mu_{R}$. This gives rise to an
applied constant external voltage $V = (\mu_{L} -\mu_{R})/e$, where
$e$ is the charge of an electron.   We restrict our study to the
case in which the electrodes are held at the same temperature. This
condition is very important in order to apply standard bosonization
techniques \cite{stone94_bosonization_book,gogolin98_1D_book}.
Indeed, as was recently explained in Ref.
\onlinecite{gutman10_bosonization_nonequilibrium}, standard
bosonization is expected to work well only in special situations
corresponding to small deviations from equilibrium.  Under this
condition the background current flowing in the wire is $I_{0} =
e^{2}V/h$ \cite{Maslov-Stone,Safi-Schulz}. If a point like impurity
is switched on in the wire at finite time, a contribution additional
to $I_{0}$ due the effect of this impurity appears and the total
current now is $I = I_{0} - I_{bs}(t)$.  In order to compute
$I_{bs}$, after employing the usual bosonization technique
\cite{Chamon-etal,Makogon-etal}, the system is described by the
following Lagrangian density:
\begin{equation}
\mathcal{L} = \mathcal{L}_{0} + \mathcal{L}_\text{imp}.
\end{equation}
The first term describes the dynamics of the conduction electrons and is modeled by a spinless Tomonaga-Luttinger liquid with renormalized Fermi velocity $v$,
\begin{equation}
\mathcal{L}_{0} = \frac{1}{2}\Phi(x,t)
\left(v^2\frac{\partial^2}{\partial x^2} -
\frac{\partial^2}{\partial t^2}\right)\Phi(x,t),
\end{equation}
where the bosonic field $\Phi$ is related to the charge density
$\Phi=e\sqrt{K}\frac{1}{2\pi}\partial_x\rho$, $v = v_{F}/K$ when $|x| < L/2$ and $v = v_{F}$ when  $|x| > L/2$, where $v_{F}$ is the Fermi velocity and $K$ represents the strength of the
electron-electron interactions. For repulsive interactions $K < 1$,
and for noninteracting electrons $K = 1$. The second term contains
the impurity contribution producing backscattering of incident waves
at the point $x_{0}$,
\begin{multline}\label{impurity lagrangian}
\mathcal{L}_\text{imp}= - \frac{g_{b}}{\pi \hbar \Lambda} \delta(x - x_{0}) W(t)\\
\times \cos\left[2 k_\text{F}x/\hbar + 2\sqrt{\pi K v}\Phi(x, t) + e V t/\hbar\right],
\end{multline}
where the function $W(t)$ models the way in which the barrier is
switched on. In this work we restrict our analysis to the case of a
sudden switch that takes place at $t=t_0$: $W(t)=\Theta (t -
t_{0})$. $\Lambda$ is a short-distance cutoff. We only consider
backscattering between electrons and impurities, since at the lowest
order in the perturbative expansion in the coupling $g_b$ forward
scattering does not change the transport properties studied here.

\begin{figure}
\begin{center}
\includegraphics{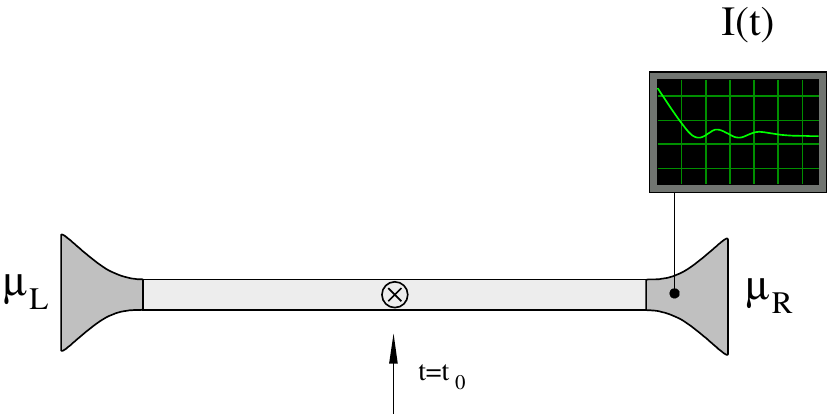}
\caption{\label{fig2ne}(Color online) The figure shows a quantum
wire coupled adiabatically to two reservoirs with different chemical
potentials, with a backscattering impurity switched on at time
$t=t_0$. The current $I(t)$ is measured as a function of time.}
\end{center}
\end{figure}

The backscattered contribution to the
current at any time $t$ is given by
\begin{equation}
I_{bs}(t) = \langle 0\vert S(- \infty ; t)\widehat{I}_{bs}(t)S(t ; -
\infty) \vert 0 \rangle.     \,   \label{uno}
\end{equation}
Here $ \vert 0 \rangle$ denotes the initial state and $S$ is the scattering matrix, which to
the lowest order in the coupling $g_{b}$ is
\begin{equation}
S(t ; - \infty) = 1 - i \int^{\infty}_{-\infty} d x \int^{t}_{- \infty}
\mathcal{L}_\text{imp} (t') d t'\,\label{dos}.
\end{equation}
$\widehat{I}_{bs}(t)$ is the operator associated to the
backscattered current which measures the rate of change of the total
number of right movers in the system due to the backscattering
impurity~\cite{feldman03_bs_impurity,sharma03_adaibatic_transport,agarwal07_transport_LL_pumping,
salvay09_pumping_LL_two_impurities},
\begin{equation}
\widehat{I}_{bs}(t) = \frac{g_{b} e}{\pi \hbar \Lambda} \Theta (t -
t_{0}) \sin\left[\frac{2 k_{F}x_{0}}{\hbar} + 2\sqrt{\pi K
v}\widehat{\Phi}(x_{0}, t) + \frac{e V t}{\hbar}\right].
\end{equation}

To avoid confusion, notice that $\widehat{I}_{bs}(t)$ is, by
definition, independent of the position on the wire $x$. Indeed,
since it is connected with the time evolution of the total number of
right moving particles, it involves an integral of the corresponding
density over the $x$-variable. Of course, it does depend on the
position of the impurity $x_0$, due to the ultralocal impurity
Hamiltonian derived from (\ref{impurity lagrangian}). In order to
compute $I_{bs}$ one substitutes Eq. (\ref{dos}) into Eq.
(\ref{uno}) and finds a series of contributions of the form
\begin{equation}
F(t - t') =\langle 0 \vert \exp[ 2 i  \sqrt{\pi K v} \widehat{\Phi}(x_{0}, t')]\exp[- 2 i  \sqrt{\pi K v} \widehat{\Phi}(x_{0}, t)] \vert 0 \rangle.\label{factorF}
\end{equation}

In order to explicitly evaluate the average appearing in  Eq.
(\ref{factorF}) we employ the Keldysh formalism, a well known
technique that renders the study of nonequilibrium systems more
accessible
\cite{lifshitz80_nonequilibrium,mahan07_book,das97_nonequilibrium,kamenev09_keldysh_review}.
We use Baker-Haussdorff formula and the fact that the commutator of
the $\widehat{\Phi}$-fields is a c-number. This allows to write the
v.e.v. of an exponential as the exponential of a v.e.v.. The result
is an expression that depends on Keldysh lesser function $i
G^{<}(x,t;x,t')=\langle 0|\widehat{\Phi}(x,t')\,\widehat{\Phi}(x,
t)| 0 \rangle$. This function depends, of course, on the size of the
system \cite{mattson97_LL_impurity_finite_T_L}, acquiring a simple form for $L \rightarrow
\infty$. In the next section, in order to make the interpretation of
the new, finite size effects, more transparent, we will briefly
recall the main results obtained in Ref.
\onlinecite{salvay10_transient_impurity} for an infinite wire.

\section{Results for $L \to \infty$}\label{sec:review}

   Restricting our analysis to the zero temperature limit (See Ref. \onlinecite{salvay10_transient_impurity} for the extension to
finite temperatures), (\ref{factorF}) reduces to
\begin{equation}
F(t - t') =  \frac{ \Lambda^{2 K} \exp[i \pi K \sign [t - t']]}{\vert v (t - t')\vert^{2 K}} \, , \label{tresa}
\end{equation}where $\sign$ is the Sign function.  First of all, we compute (\ref{uno})  with
the impurity switched on at time $- \infty$ :

\begin{equation} I_{bs}(\infty) = \frac{g_{B}^{2} e  \Lambda ^{2 K - 2}}{2 \pi \hbar^{2} v ^{2 K}\Gamma[2 K]}\left|\frac{e V}{\hbar}\right|^{2 K -1} \sign[V].\label{pu}\end{equation}
This corresponds to the well-known case of a static
impurity\cite{kane92_fisher_resonant_tunneling_LL}, where the backscattered current goes as $V^{2 K
- 1}$. We note that for $K < 1/2$, the backscattered current becomes
large when $V$ decreases. Hence, the perturbative expansion in
powers of $g_{B}$ breaks down when $V \rightarrow 0$.  Using a
scaling analysis we can estimate that this expansion is valid when
$\frac{g_{B}}{\hbar v} (\frac{\Lambda e V}{ \hbar v})^{K - 1} \ll
1$. We emphasize that expression (\ref{pu}) does not include the
case $V = 0$, where the current is zero too.  All these statements
imply that the current must be a nonmonotonic function of $V$.  In
order to determine this function one has to go beyond the
lowest-order perturbative results of this work.

When the impurity is switched on at a finite time $t_{0}$, the
current exhibits for finite times a relaxation dynamics to its
asymptotic value $I_{bs}(\infty)$. The time-dependent backscattering
current dynamics acquires a complicated form, yet it can be written
in terms of known analytic functions:\begin{multline} I_{bs}(t -
t_{0}) = \frac{\Theta(t - t_{0})\Gamma[2 K] [\omega_{0} (t - t_{0})]
^{2 - 2 K}}{\Gamma[K]\Gamma[2 - K]} \\ \times {}_{1}F_{2}( 1 - K ;
3/2 , 2 - K ; - [\omega_{0} (t - t_{0})/2]^{2})\,I_{bs}(\infty)
,\label{current}
\end{multline}
where $_{1}F_{2}$ is the generalized hypergeometric function and  $\omega_{0} = \frac{e |V| }{\hbar}$  is the Josephson frequency related to the source voltage. This is
a damped oscillatory function of $\omega_{0} (t - t_{0})$ with period $2 \pi$ and
relative maxima in $\omega_{0} (t - t_{0}) = (2 n + 1) \pi$ with $n$ natural. In the long times regime $\omega_0(t-t_0)\gg 1$, this expression can be approximated by its asymptotic form
\begin{multline}
I_{bs}(t - t_{0}) \approx  \Theta(t - t_{0})\{1 \\ - \frac{ [\omega_{0} (t - t_{0})] ^{- 2 K}}{\cos[\pi K] \Gamma[1 - 2
K]}\cos[\omega_{0} (t - t_{0})] \}I_{bs}(\infty) \label{g}.
\end{multline}

Since the change in the backscattered current due to the sudden switching
behaves as $(t - t_{0}) ^{- 2 K}$, we conclude that the relaxation of the
system is faster for small electron-electron interaction (For the
sake of clarity, let us stress that this relaxation characterizes
the transition of the backscattering current between two
off-equilibrium regimes). We thus find an explicit connection
between electron interactions and the switching time of the
externally controlled barrier: the stronger the correlations, the
longer the persistence of the non adiabatic effect.  A
determination of the current as function of time with a temporal
resolution smaller than $\frac{2 \pi}{\omega_{0}}$ is a direct method
to obtain the exponent of the temporal decay, and then, the $K$
value of the quantum wire. We emphasize that this proposed method is
performed at constant source-drain voltage.

\section{Finite size and switching effects}

\begin{figure}
\begin{center}
\includegraphics[width=\figwidth]{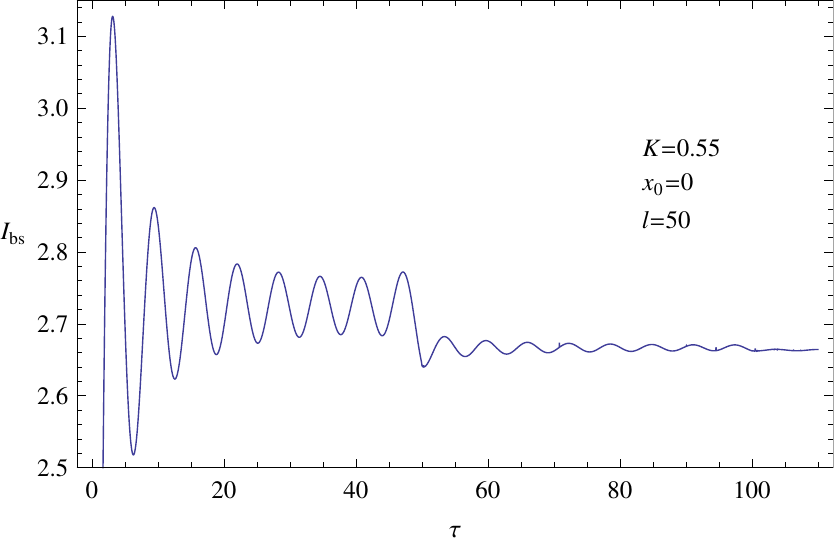}
\caption{Typical behavior of $I_{bs}$ in units of $I_{bs}(\infty)$  for $\ell \gg 1$
and the impurity located at the center of the wire.}
\end{center}
\end{figure}

\begin{figure}
\begin{center}
\includegraphics[width=\figwidth]{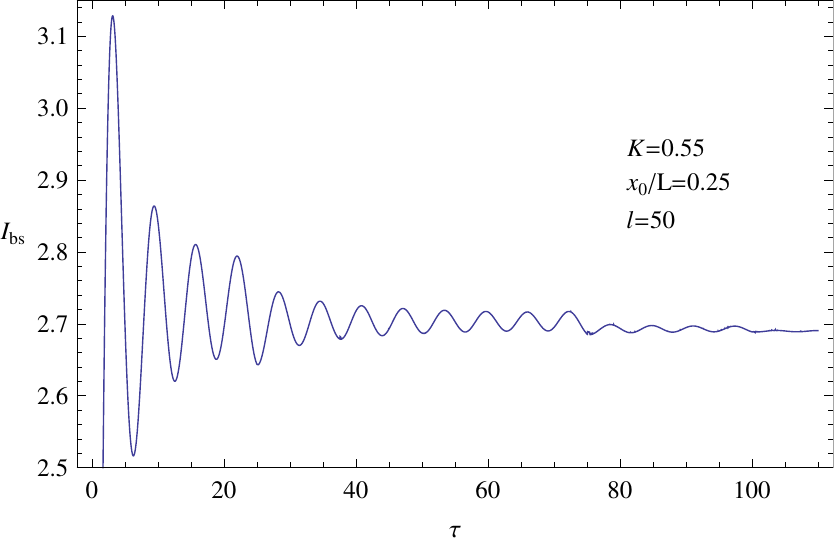}
\caption{Typical behavior of $I_{bs}$ in units of $I_{bs}(\infty)$  for $\ell \gg 1$
and the impurity located at $x_0/L=0.25$.}
\end{center}
\end{figure}

In this section we present the main results of this work. We will
show how the combined effects of a sudden switch-on of an externally
controlled barrier and the quantum interference originated by
reflections at the ends of the wire (which are now at finite
distances from the barrier), modify the time evolution of the
backscattered current. The crucial point is that, when the wire has
a finite length $L$ the function (\ref{factorF}) becomes dependent
on $L$ and $x_0$. In this case we obtain\begin{multline}
 F(t - t') = \prod_{n \text{ even}} \left[  \frac{ (\delta + i \omega_{B} (t - t'))^{2} + n^{2}}{\delta^{2} + n^{2}} \right]^{- K\gamma^{\vert n\vert}}\\ \times  \prod_{n \text{ odd}} \left[  \frac{ (\delta + i \omega_{B} (t - t'))^{2} + (n - 2 x_{0}/L)^{2}}{\delta^{2} + (n - 2 x_{0}/L)^{2}} \right]^{-K\gamma^{\vert n\vert}}.\label{tres}
\end{multline}
where $\gamma = (1 - K)/(1 + K)$ is the Andreev-like reflection
parameter, $\omega_{B} = v_{F}/KL$ is the ballistic frequency related to
the system length and $\delta=\Lambda/L$. In view of this result,
from now on we shall write $F(t)\equiv F(\omega_{B} t)$. At this
point, replacing (\ref{tres}) in (\ref{uno}), we find the following
expression for $I_{bs}$:
\begin{equation}
I_{bs}(t - t_{0}) = C\Theta(t - t_{0})\int^{t - t_{0}}_{0}dt'
\sin(\omega_{0} t' )\Im F(\omega_{B} t')\label{tresd},
\end{equation}
where $C=\frac{-e g_{B}^{2}\sign[V]}{\pi^{2} \hbar^{2}\Lambda^2}$.
This is the generalization, for finite $L$, of the result first
obtained in Ref. (\onlinecite{salvay10_transient_impurity}) for an
infinitely long wire. As a first test of the validity of this
expression we have checked that taking the limit $L \rightarrow
\infty$ one recovers the result given in Eq.(\ref{current}). In
contrast to that case, in the present situation the derivation of an
approximate expression that could facilitate the physical
interpretation is not a trivial task. Despite of this fact, it is
possible to obtain an analytical expression for (\ref{tresd})
following the lines explained in Ref.
(\onlinecite{ponomarenko97_asymptotic_expansion}). Let us first
define the dimensionless parameter $\ell = \omega_{0}/\omega_{B}$ so
that for $ \ell \gg (\ll) 1$ we have the case of large (small)
length. (For example, for applied voltages of order $1 mV$ and $1
\mu V$, this corresponds to lengths of order $L \gg (\ll) 10^{-6} m$
and $L \gg (\ll) 10^{-3} m$, respectively). For $\ell \gg 1$ one can
get an asymptotic expansion for the current, taking into account
that the function $F(\omega_{B} t)$ has an infinite series of cuts
in the complex z-plane ($z=\omega_{B} t$), going from $z=i\infty \pm
2 n$ to $z=i\delta \pm 2 n$ and from $z=i\infty \pm (2
n+1)(1+2x_0/L)$ to $z=i\delta \pm (2 n+1)(1+2x_0/L)$, with $n$
integer. The integral can be cast as a sum of contours going around
these cuts. The result is a lengthy expression. For the sake of
clarity, we will display here the result for the symmetrical case
($x_0=0$), which contains the main ingredients needed for a physical
interpretation and at the same time acquires a more compact form:
\begin{widetext}
\begin{multline}\label{analyt}
I_{bs}(\tau)=-\Theta(\tau)I_{bs}(\infty)\Im\Bigg\{2 \Gamma[2 K]\sin[\pi K]\frac{\tau^{1-2K}e^{-i\tau}}{\pi(1-2K)}\Phi(1,2-2K,i\tau)\\
+\sum_{n=1}^{\infty}\frac{D_{n}\,\ell^{2K(\gamma^{n}-1)}}{\pi(1-2K\gamma^{n})}2 \sin(\pi K) \Gamma[2 K] \Big[(nl+\tau)^{1-2K\gamma^{n}}e^{-i\tau}\Phi(1,2-2K\gamma^{n},i(nl+\tau)) \\ -(nl-\tau)^{1-2K\gamma^{n}}e^{+i\tau}\Phi(1,2-2K\gamma^{n},i(nl-\tau))\Big]\\
+\sum_{n=1}^{\infty}\frac{D_{n}\,\ell^{2K(\gamma^{n}-1)}}{\pi(1-2K\gamma^{n})}\Theta(\tau-n\ell)2 \sin(2\pi K\gamma^{n})\Gamma[2 K] e^{i\pi K}e^{-i\tau}(\tau-nl)^{1-2K\gamma^{n}}\Phi(1,2-2K\gamma^{n},i(\tau-nl))\Bigg\}
\end{multline}
\end{widetext}
where $\Phi$ is the Kummer confluent hypergeometric function and we have defined $\tau = \omega_{0}(t - t_{0})$
and the numerical coefficient:
\begin{equation}D_{n} = 2^{- 2 K \gamma^{n}}n^{2 K(\gamma^{n} - 1)}\prod^{\infty}_{m \neq n, m > 0} |\frac{m^{2}}{m^{2} - n^{2}}|^{2 K \gamma^{m}}.\end{equation}

Equation (\ref{analyt}) turns out to be a valuable tool in order to
understand the mechanisms that govern the transient process of
$I_{bs}$, even in the general case ($x_0=0$).

In Figures 2 and 3 we display the time evolution of $I_{bs}$ as a
function of $\omega_{0} t$ (we set $t_{0}=0$) for long wires ($\ell
\gg  1$), with the barrier at the center ($x_{0} = 0$) and displaced
($x_{0}/L = 0.25$). The current evolves through ``temporal steps" that
take place in $\tau$-regions with ends at $\tau = 2 n \ell$ and
$\tau = (2 n + 1 \pm 2 x_{0}/L)\ell$. For $x_{0} = 0$ the occurrence
of these steps becomes apparent through the functions $\Theta(\tau -
n \ell)$ appearing in the analytical expression (\ref{analyt}), in
which each term represents a new step. One can deduce an expression
that describes these steps by retaining only the long times
asymptotics in Eq. (\ref{analyt}):
\begin{widetext}
\begin{equation}\label{eq:steps}
I_{bs}^{steps}(\tau)=\Theta(\tau)I_{bs}(\infty)\left\{1+2\sum_{n=1}^{\infty}D_{n}\frac{\Gamma(2K)\cos[nl-\pi K(1+\gamma^n)]}{\Gamma(2K\gamma^n)\ell^{2K(1-\gamma^{n})}}\Theta(\tau-n\ell)\right\}
\end{equation}
\end{widetext}
In the limit $\tau\to\infty$ Eq. (\ref{eq:steps}) reduces to the
static value of the current. In this last case, the current is a
damped oscillatory function of $\ell$ with period $2 \pi$ and the
dominant exponent in the decay is $2 K(1 - \gamma)$, i.e the current
has the form $I_{bs} \approx I_{bs}(\infty)(1 + C_{0}\cos[\ell - \pi
K ( 1 + \gamma)]\ell^{- 2 K(1 - \gamma)})$, where $C_{0}$ is a
coefficient that depends of $K$.  When $x_{0} \neq 0$, $I_{bs}$ is a
superposition  of two damped oscillatory functions of $\ell$ with
period $2 \pi/( 1 \pm 2 x_{0}/L)$ and the dominant exponent in the
decay is $2 K(1 - \gamma/2)$ (the same for both functions).

In addition to (\ref{eq:steps}), Eq. (\ref{analyt}) exhibits an oscillatory behavior in $\tau$ with period equal to $2 \pi$ and
maxima located at $\tau  = (2 m + 1)\pi$ as in the infinite length case. Those $\tau$-regions are
more pronounced and become more visible for $K \ll 1$, i.e.
transitory effects originated in the abrupt switching are more
important for stronger correlations between electrons. For very long
times the current approaches the value corresponding to the case in
which one has a static impurity immersed in a wire of length $L$.

Now, we analyze the envelope of the oscillatory function found for
the backscattered current, for the case of a long wire ($ l \gg 1$).
First of all one observes that, in contrast to the case
$L\rightarrow\infty$, this envelope is not a monotonic function
anymore. Within each $\tau$-window (letting aside, of course, the
first abrupt growth at $t=0$) there is an initial decay, followed by
a revival of the current that lasts until the beginning of the next
step. Again, when the impurity is at the center of the wire (See
Fig. 2), the power laws that govern the evolution of $I_{bs}$ can be
read from (\ref{analyt}). In each of these initial relaxation
processes the envelope of the current behaves as $(\tau - n\ell)^{-
2 K \gamma^{n}}$, for $1+n\ell<\tau \ll (n+1)\ell$, $n$ being a
natural number that labels the windows. In particular, at the
beginning of the first temporal step that corresponds to $n=0$ (that
is, the ``short time" region $(\tau \ll \ell)$), the decay is the
same as in the case of an infinite wire, found in (\ref{g}):
$\tau^{- 2 K}$.

For an asymmetrical arrangement in which $x_{0} \neq 0$, the general
pattern of the evolution is distorted (See Fig. 3). In particular
the heights of the temporal steps (the values of the current
envelope in each step) do not decrease monotonically, and their
durations are not all equal (they were all equal to $\ell$ when
$x_{0}=0$). Moreover, these durations follow an alternated pattern
of the form $\Delta\tau_1$, $\Delta\tau_2$, $\Delta\tau_1$, with
$\Delta\tau_1=(1 - 2 x_{0}/L)\ell$ and $\Delta\tau_2=(4
x_{0}/L)\ell$ (Note that $2\Delta\tau_1+\Delta\tau_2=2\ell$). The
decay laws that we found for these $\tau$-windows are:
$$
(\tau - 2p\ell)^{- 2 K\gamma^{2p}}
$$
for $1+2p\ell<\tau \ll 2p\ell+\ell(1 - 2 x_{0}/L)$;
$$
(\tau - 2p\ell-\ell(1 - 2 x_{0}/L))^{- K\gamma^{2p+1}}
$$
for $1+2p\ell+\ell(1 - 2 x_{0}/L)<\tau \ll 2p\ell+\ell(1 + 2
x_{0}/L)$ and
$$
(\tau - 2p\ell-\ell(1 + 2 x_{0}/L))^{- K\gamma^{2p+1}}
$$
for $1+2p\ell+\ell(1 + 2 x_{0}/L)<\tau \ll (2p+2)\ell$. Here the
natural number $p$ does not label windows. The value $p=0$ gives the
behavior corresponding to the first three $\tau$-windows that appear
in the interval $0<\tau<2\ell$. This basic pattern is repeated along
the subsequent intervals of duration $2\ell$, the corresponding
exponents are given by the following values of $p$.

A word of caution is in order here regarding the limit $x_0\rightarrow0$. Notice that when considering this limit in the expressions above one does not
recover the exponents for an
impurity placed at the center of the wire. This is so because the correlation function corresponding to the general case ($x_{0} \neq 0$) has pairs of different singularities that merge
when $x_{0}=0$. A similar situation is discussed in the study of finite length effects for an static impurity presented in Ref.\onlinecite{dolcini05_finite_length_temperature}.

The relationship between the relaxation process and the strength of Coulomb electron correlations revealed in our analysis provides an alternative way to determine the Luttinger parameter $K$ through
time measurements. Since the total current has the form $I_0 - I_{bs}(t)$, a determination of the current as a function of time (keeping the  source-drain voltage constant) with a temporal
resolution smaller than $2\pi\hbar/eV$, enables to obtain the exponent of the temporal decay, and then the $K$ value of the
quantum wire. Interestingly, having taken into account the finite
size of the system, our results suggest another possible
experimental application. Indeed, the experimental determination of
the duration of the $\tau$-windows described above allows to obtain
the value of the ratio of Josephson to ballistic frequencies $\ell=\frac{eVL}{\hbar}\frac{K}{v_{F}}$. Thus, if $V$, $L$
and the Fermi velocity $v_{F}$ are known (the last one can be obtained
by photoemission spectroscopy), this simple technique gives another
way of finding the Luttinger parameter.

Finally, for completeness, we explore transport properties in short
wires ($\ell < 1$). In this case the asymptotic expansion described
above is not valid and then we are led to analyze Eq. (\ref{tresd})
numerically. In Figure 4 we show the behavior of $I_{bs}$ for short
wires . In this case the structure of ``temporal steps" disappears
and finite-time switching effects are relevant for $\omega_{0}t \leq
1$. In this regime one observes a monotonous growth of the
backscattered current, with discontinuities in the time-derivative
that take place at $\omega_{0}t = 2 n \ell$ and $\omega_{0}t = (2 n
+ 1 \pm 2 x_{0}/L)\ell$. For $\omega_{0}t > 1$ the current also
tends to the static-impurity value $I_{bs}(\infty)$, as expected.
Thus, in practice, for short wires, the oscillation of the current
as function of $\omega_{0}t$ ceases to be appreciable. This is due
to the fact that, for fixed $K$ and $V$, the Josephson current
$\omega_{0}$ becomes negligible when compared with the ballistic
frequency $\omega_{B}$. In contrast to what happens for long wires,
where an ultrafast current growth is followed by an oscillatory
decay pattern, here $I_{bs}$ undergoes a rapid increase towards the
steady-state current, which goes as $\ell^{2 - 2K}$ and has a
monotonic decrease with $|x_{0}/L|$. In this sense one could say
that, in a sufficiently short wire, instead of a relaxation process
(after the initial jump), a monotonic enhancement, saturating at the
steady-state current, is predicted.
\begin{figure}
\begin{center}
\includegraphics[width=\figwidth]{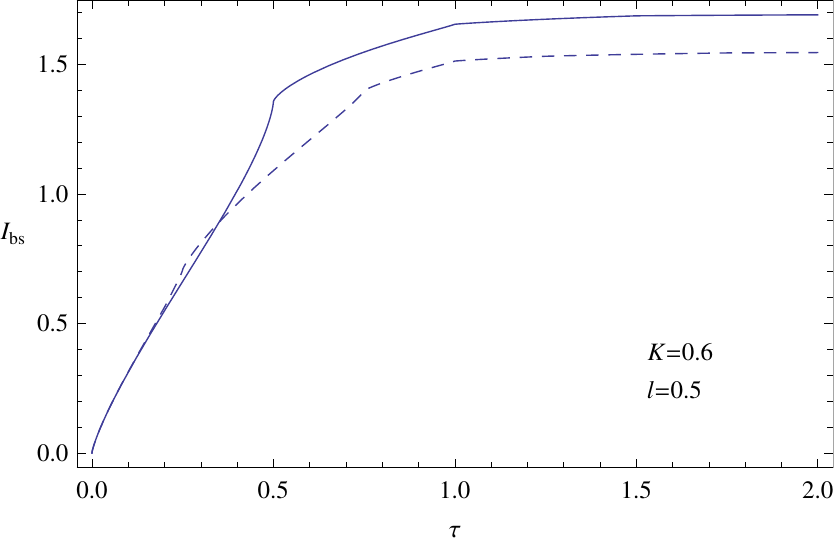}
\caption{Typical behavior of $I_{bs}$ in units of $I_{bs}(\infty)$ for $\ell < 1$
and the impurity located at $x_0 =0 $ (solid line) and $x_0/L = 0.25$
(dashed line).}
\end{center}
\end{figure}

\section{Conclusions}

To summarize, we have analyzed the characteristics of time-dependent
transport in a Tomonaga-Luttinger liquid subjected to a constant
bias voltage, when a weak barrier is suddenly switched on. The novel
features of our investigation come from the consideration of a wire
with a finite length $L$. We focused our attention on the
backscattered current $I_{bs}$ which is originated by the abrupt
appearance of the barrier. We showed that, in comparison with the
case $L\rightarrow\infty$, for which the current relaxes to the
steady-state value with an envelope that obeys the simple law $t^{-
2K}$, finite size effects lead to a much more complex and rich
evolution. In order to grasp the main features of this dynamics, we
have identified different regimes in terms of the Josephson
frequency $\omega_{0} = \frac{e |V|}{\hbar}$ and the ballistic
frequency $\omega_{B} = \frac{v_{F} K}{L}$. In particular, when the
applied bias ($V$) and the electron-electron correlations ($K$) are
held fixed, the ratio $\ell=\omega_{0}/\omega_{B}$ provides a
natural parameter to characterize ``long" ($ \ell \gg  1$) and
``short" ($ \ell \ll  1$) wires. In this context it becomes natural
to compare the results obtained for $ \ell \gg  1$ with the
previously known behavior found in infinite systems. We found that
for long sizes the evolution of $I_{bs}$ presents a new structure of
temporal steps, each of which with a different decay law with the
generic form $(t-\nu)^{-  \mu K \gamma^{m-1}}$, where $\nu$ depends
on $m$, $\omega_{B}$ and $x_0$, and $\gamma = (1 - K)/(1 + K)$ is an
Andreev-like reflection parameter. The constant $\mu$ is simply 1 or
2, depending on the $\tau$-step one is looking at. Since $K<1$ and
$m$ is a natural number that labels each time window, being $m$
bigger for longer times, the older the window the slower the
relaxation. These results display the way in which the combined
effects of electronic correlations and quantum interference,
originated by reflected waves at the interfaces with the electrodes,
affect the transient process.

Concerning short wires, the whole structure of temporal steps
disappears and the evolution becomes simpler, showing a monotonic
growth of $I_{bs}$ towards its steady-state value.

From the point of view of the potential applications of our study it
is specially interesting the case of long wires, for which we were
able to derive approximate asymptotic expressions from which one
determines the durations of the temporal steps and the power laws
obeyed by the envelope of the current in the time intervals where a
relaxation takes place. If one measures the decay of the current
along these intervals, the value of $K$ can be obtained from the
exponents $\mu K ((1 - K)/(1 + K))^{m-1}$, where the values of both
$\mu$ and $m$ are univocally given by the $\tau$-window for which
the measurement is performed. A second method, probably easier to
implement, is also indicated by our findings. Indeed, as explained
in the text, just by measuring the endpoints of the $\tau$-windows,
taking into account that $L$ and $x_0$ are assumed to be known
(remember that in the context of our investigation the impurity is
assumed to be due to an external gate), one can determine the value
of $\ell=\frac{eVL}{\hbar}\frac{K}{v_{F}}$. Since the bias voltage $V$
is also externally controlled and held fixed, this simple technique
gives a direct way of finding the ratio $\frac{K}{v_{F}}$. Of course, if
one of these two parameters are known by means of another reliable
method, the remaining one is directly obtained from our recipe.

\vspace{1cm}

\acknowledgments  This work was partially supported by Universidad
Nacional de La Plata (Argentina) and Consejo Nacional de
Investigaciones Cient\'{\i}ficas y T\'ecnicas, CONICET (Argentina).
The authors are grateful to Fabrizio Dolcini, Hugo Aita and Pablo
Pisani for helpful discussions.

%\bibliography{impurity,phys}

\end{document}